\documentclass[a4paper,twocolumn,11pt,authblk,unpublished]{quantumarticle}
\pdfoutput=1
\usepackage[utf8]{inputenc}
\usepackage[english]{babel}
\usepackage[T1]{fontenc}
\usepackage{amsmath}
\usepackage{amssymb}
\usepackage{hyperref}
\usepackage{enumitem}
\usepackage{booktabs}
\usepackage{tikz}
\usepackage{lipsum}
\usepackage[numbers]{natbib}
\usepackage{braket}       % 支持 \bra{}, \ket{}, \braket{} 语法
\usepackage{mathtools}    % 支持 \xrightarrow 等增强数学命令

\begin{document}

\title{Co-designed Quantum Discrete Adiabatic Linear System Solver Via Dynamic Circuits}

\author[1]{Boxuan Ai}
\author[1,3]{Shuo He}
\author[2]{Xiang Zhao}
\author[2]{Lin Yang}
\author[2]{Guozhen Liu}
\author[1]{Pengfei Gao}
\author[1]{Hongbao Liu}
\author[1]{Tao Tang}
\author[2]{Jiecheng Yang}
\email{yangjiecheng@turingq.com}
\author[3]{Jie Wu}
\email{jwu@fudan.edu.cn}

%--- 按你要的顺序定义机构 ---%
\affil[1]{China Unionpay, Shanghai, China}
\affil[2]{TuringQ Co. Ltd., Shanghai, China}
\affil[3]{Fudan University, Shanghai, China}

\maketitle

\begin{abstract}
Existing quantum discrete adiabatic approaches are hindered by circuit depth that increases linearly with the number of evolution steps, a significant challenge for current quantum hardware with limited coherence times. To address this, we propose a co-designed framework that synergistically integrates dynamic circuit capabilities with real-time classical processing. This framework reformulates the quantum adiabatic evolution into discrete, dynamically adjustable segments. The unitary operator for each segment is optimized on-the-fly using classical computation, and circuit multiplexing techniques are leveraged to reduce the overall circuit depth scaling from \(O(\text{steps}\times\text{depth}(U))\) to \(O(\text{depth}(U))\). We implement and benchmark a quantum discrete adiabatic linear solver based on this framework for linear systems of \(W \in {2,4,8,16}\) dimensions with condition numbers \(\kappa \in {10,20,30,40,50}\). Our solver successfully overcomes previous depth limitations, maintaining over \(80\%\) solution fidelity even under realistic noise models. Key algorithmic optimizations contributing to this performance include a first-order approximation of the discrete evolution operator, a tailored dynamic circuit design exploiting real-imaginary component separation, and noise-resilient post-processing techniques.
\end{abstract}

\section{Introduction}
Efforts to realize quantum advantage in solving linear systems have catalyzed significant advancements in quantum algorithm development. Originating from the seminal Harrow-Hassidim-Lloyd (HHL) algorithm \cite{Harrow2008QuantumAF}, which first demonstrated exponential speedups for sparse linear systems, the field has further evolved through two pivotal methodologies: the Variational Quantum Linear Solver (VQLS) and Adiabatic Quantum Computing (AQC) \cite{BravoPrieto2019VariationalQL, Costa2021OptimalSQ,Suba2018QuantumAF,An2019QuantumLS,Lin2019OptimalPB,Costa2023TheDA,cao2013quantum,chakraborty2018power,childs2017quantum,gilyen2019quantum,wossnig2018quantum,xu2021variational,jennings2024cost}. Table 1 summarizes key milestones in complexity reduction.

The VQLS approach leverages parameterized quantum circuits to approximate solutions efficiently via shallow circuit depths. Nevertheless, its practical effectiveness heavily depends on classical optimization routines and is often constrained by the barren plateau phenomenon, leading to vanishing gradients \cite{McClean2018BarrenPI, Cerezo2021CostFD, Wang2020NoiseinducedBP}. Conversely, adiabatic methods theoretically guarantee convergence by gradually evolving a Hamiltonian from an easily prepared initial state toward a target solution state. However, traditional adiabatic techniques require circuit depths that scale as  relative to the desired precision  \cite{Costa2021OptimalSQ}, primarily due to the sequential application of time-dependent unitary operations. Such requirements severely restrict scalability on current quantum hardware with limited coherence times.

\begin{table*}[htb]
\centering
\begin{tabular}{@{}lllll@{}}
\toprule
\textbf{Year} & \textbf{Reference} & \textbf{Primary innovation} & \textbf{Complexity} \\ \midrule
2008 & Harrow, Hassidim, Lloyd \cite{Harrow2008QuantumAF}& Phase estimation & $O(\kappa^2 / \epsilon)$ \\
2012 & Ambainis \cite{ambainis2012variable}& Variable-time amplitude amplification & $O(\kappa (\log(\kappa) / \epsilon)^3)$ \\
2018 & Subasi, Somma, Orsucci \cite{Suba2018QuantumAF}& Adiabatic randomization method & $O((\kappa \log \kappa) / \epsilon)$ \\
2019 & Lin, Tong \cite{Lin2019OptimalPB}& Adiabatic plus eigenstate filtering & $O(\kappa \log(\kappa / \epsilon))$ \\
2021 & Pedro C. S. Costa \cite{Costa2023TheDA}& Discrete adiabatic theorem & $O(\kappa \log(1 / \epsilon))$ \\ \bottomrule
\end{tabular}
\caption{Summary of advancements in quantum algorithms. \( \kappa \) represents the condition number of matrix \( A \), while \( \epsilon \) denotes the desired level of precision for preparing the state \( x \).}
\end{table*}

Recent progress in dynamic quantum circuits \cite{corcoles2021exploiting, baumer2024efficient, CarreraVazquez2024CombiningQP,yu2024symmetry,fang2023dynamic} offers a promising avenue to address these scalability issues. Notably, IBM Quantum’s breakthrough in 2024 \cite{CarreraVazquez2024CombiningQP} demonstrated that mid-circuit measurements combined with real-time adaptive control significantly enhance the scale and connectivity of quantum processors, providing a novel mechanism to overcome critical depth constraints in quantum algorithm implementations.

Leveraging these developments, we identify a fundamental limitation in existing quantum discrete adiabatic methods \cite{Costa2021OptimalSQ, Lin2019OptimalPB, Costa2023TheDA, Suba2018QuantumAF, An2019QuantumLS}, namely, the linear scaling of circuit depth with the number of evolution steps. This scaling directly conflicts with the limited coherent time of current quantum hardware. To address this issue, we propose a co-designed framework that integrates dynamic circuit capabilities with classical optimization. Specifically, we partition the quantum adiabatic evolution into discrete segments, dynamically adjusting and optimizing the unitary operations of each segment concurrently via classical computation (see Fig. 1). By using measurement - assisted computation, we can split a deep quantum circuit into several shallower ones. This improves the fidelity of the solution within the qubit decoherence time. Details are provided in Section 3. This approach, combined with circuit multiplexing, reduces depth complexity from $O(\text{steps}\times \text{depth}(U))$ to $O(\text{depth}(U))$, significantly curtailing the overall circuit depth \cite{Costa2021OptimalSQ,Costa2023TheDA}.

\begin{figure*}[htb]
	\centering
	\includegraphics[width=\linewidth]{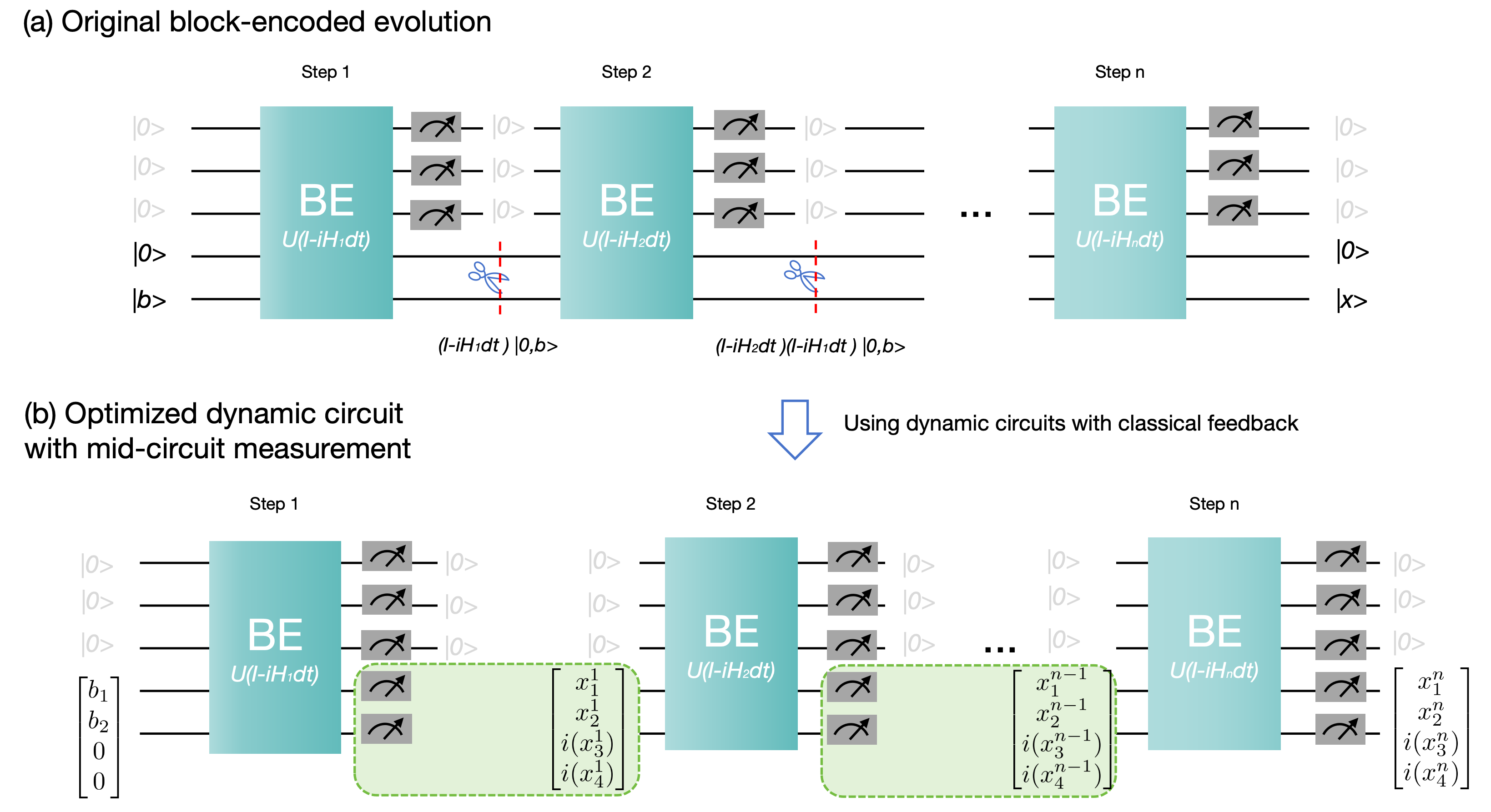}
	\caption{Comparison of quantum circuits for the discrete adiabatic solver. 
	(a) The original circuit implemented via block encoding, with a depth of $O(\text{steps} \times \text{depth}(U))$. 
	(b) The optimized circuit proposed in this work leverages dynamic circuits to reduce the depth to $O(\text{depth}(U))$.}
	\label{fig:be-circuits}
\end{figure*}

This work presents the pioneering implementation of a quantum discrete adiabatic linear solver explicitly optimized for depth constraints through dynamic circuits. Through benchmarking on various linear systems with different condition numbers, our approach consistently achieves solution fidelities exceeding 80\%, even under realistic noise conditions. This co-design methodology, integrating algorithmic structure with dynamic circuit capabilities, offers a novel strategy for constructing resource-efficient quantum linear solvers suitable for the Noisy Intermediate-Scale Quantum (NISQ) era\cite{Preskill2018QuantumCI}.

The rest of the paper is structured as follows. Section 2 revisits the theoretical foundation underlying the quantum discrete adiabatic solver \cite{Costa2021OptimalSQ}, including Hamiltonian construction, adiabatic interpolation, and discretized evolution. Section 3 elaborates on our algorithmic enhancements, comprising first-order approximations of the evolution operator, dynamic circuit design with real-imaginary component separation, and fidelity enhancement through post-processing techniques. Section 4 presents detailed numerical analyses assessing algorithm performance across varied system sizes , noise models and noiseless models. Finally, we then conclude in Section 5.

\section{Quantum Discrete Adiabatic Linear System Solver}
\subsection{Adiabatic Quantum Computing}
The quantum adiabatic theorem, which serves as a foundational theory in quantum computing, is rooted in the essential aspects of quantum dynamics. For a quantum system with a Hamiltonian $H(t)$ dependent on time, the instantaneous eigenstates are characterized by $H(t)|n(t)\rangle=E_n(t)|n(t)\rangle$. Assuming the system starts in the ground state $|0(0)\rangle$, the theorem asserts that the system will persist in the instantaneous ground state with a probability close to one during the evolution time $T$, as long as the following inequality is satisfied \cite{albash2018adiabatic,mori2024experimentally}:
\begin{equation}
    \max _{t \in[0, T]} \frac{|\langle 1(t)| \dot{H}(t)| 0(t)\rangle \mid}{\left[E_1(t)-E_0(t)\right]^2} \ll 1
\end{equation}
Here, $E_1(t)-E_0(t)$ denotes the instantaneous energy gap.
In practical executions of adiabatic algorithms, the continuum of the evolution interval $[0, T]$ is segmented into $L$ equal divisions, each with a duration of $dt = T / L$. In this context, the time-dependent Hamiltonian is approximated as a piecewise-constant sequence $ H_k = H(k \cdot dt) $, and the total evolution operator is constructed via a time-slicing approximation:

\begin{equation}
    U(T) \approx \prod_{k=1}^L e^{-i H_k dt}
\end{equation}
For maintaining fidelity, the discretization process must adhere to the stepwise validity condition $\| H_k - H_{k-1} \| \cdot dt \ll 1$, ensuring that the discrete evolution remains an accurate approximation of the continuous adiabatic process. The discrete algorithm focuses on optimizing both step size allocation and the overall evolution time to preserve adiabaticity, even when iterations are finite. A key difficulty lies in striking a balance between errors introduced by discretization and computational expenses, which is addressed through thorough error-bound analysis and resource-efficient scheduling of quantum operations \cite{Costa2021OptimalSQ,Costa2023TheDA}.

\subsection{Quanutm Linear System Problem}
The problem of quantum linear systems (QLSP) involves solving linear equations expressed as $A {x}={b}$, where $A$ is an $N \times N$ real matrix, and ${b}$ is a vector. In the context of quantum computing, the aim is to prepare a quantum state $|x\rangle$ that is proportional to the solution vector ${x}$.

\subsubsection{Adiabatic Framework}
Initially, the quantum system is configured in the ground state of a straightforwardly constructible Hamiltonian $H_0$. Following this, the system's Hamiltonian undergoes an adiabatic evolution from $H_0$ to $H_1$, with the ground state of $H_1$ representing the encoded solution. The Hamiltonian during this evolution is expressed as:
\begin{equation}
    H(s)=[1-f(s)] H_0+f(s) H_1
\end{equation}
where $s$ ranges continuously from 0 to 1 while $f(s)$ similarly progresses from 0 to 1. $s$ changes slowly and gradually to ensure the system undergoes an adiabatic evolution.

Upon finalization, the system estimates the ground state of $H_1$, resulting in the state $|x\rangle$ that represents the solution. Subsequently, quantum measurement techniques are utilized to derive the solution.

The computational complexity of this method is largely determined by the condition number $\kappa$ of the matrix $A$ and the desired accuracy $\epsilon$. By appropriately tuning the algorithm, it achieves a complexity of $O(\kappa \log (1 / \epsilon))$ \cite{Costa2021OptimalSQ,Costa2023TheDA}.

\subsubsection{Hamiltonian Construction}
\begin{enumerate}[label=\arabic{enumi}), leftmargin=*]
\item The construction of the initial Hamiltonian, denoted as $H_0$, necessitates that its ground state be both straightforward to prepare and pertinent to the problem at hand. In the context of the linear system $A|x\rangle = |b\rangle$, we employ the following approach \cite{Costa2021OptimalSQ,An2019QuantumLS,Costa2023TheDA}:
\begin{equation}
    H_0=\left(\begin{array}{cc}
    0 & Q_b \\
    Q_b & 0
\end{array}\right), \quad Q_b=I-|b\rangle\langle b|
\end{equation}
Here, \(Q_b\) denotes the projection operator onto the subspace orthogonal to \(|b\rangle\). The state \(|\psi_0\rangle = |0, b\rangle\) explicitly corresponds to \(|b\rangle\) and satisfies the equation \(H_0 |\psi_0\rangle = 0\) with eigenvalue 0.

\item For the target Hamiltonian \(H_1\), the development of \(H_1\) for a Hermitian and positive definite matrix \(A\) is achieved by incorporating \(A\) into a Hermitian structure \cite{Costa2021OptimalSQ,An2019QuantumLS,Costa2023TheDA}:
\begin{equation}
    H_1=\left(\begin{array}{cc}
    0 & A Q_b \\
    Q_b A & 0
    \end{array}\right)
    \end{equation}
The state $\left|\psi_1\right\rangle = |0, x\rangle$, encapsulates the solution vector corresponding to the condition $A|x\rangle = |b\rangle$, which satisfies the equation \(H_1 |\psi_1\rangle = 0\) with eigenvalue 0.

\item For a not positive definite
or Hermitian matrix $A$, we extend it to form a Hermitian matrix:
\begin{equation}
    \mathcal{A}=\left(\begin{array}{cc}
    0 & A \\
    A^{\dagger} & 0
    \end{array}\right)
\end{equation}

\item Time-dependent Hamiltonian interpolation models the adiabatic evolution of a quantum system with a family of interpolating Hamiltonians \cite{Costa2021OptimalSQ,An2019QuantumLS,Costa2023TheDA}:
\begin{equation}
    H(s)=[1-f(s)] H_0+f(s) H_1, \quad s \in[0,1].
\end{equation}
The construction of these Hamiltonians is to ensure that the quantum system can evolve adiabatically from a known and easily preparable ground state \(|0, b\rangle\) of \(H_0\) to a specific eigenstate \(|0, x\rangle\) of \(H_1\), which is associated with the solution of a linear system. This evolution can be achieved by carefully designing a time-dependent Hamiltonian \(H(s)\), where \(s\) is a parameter that varies from 0 to 1 to interpolate the evolution path between \(H_0\) and \(H_1\). The selection of $f(s)$ is crucial in determining efficiency. A linear function $f(s)=s$ meets the boundary requirements $f(0)=0$ and $f(1)=1$, while maintaining a constant derivative:
\begin{equation}
    \frac{d f}{d s}=1
\end{equation}
Although straightforward to apply, this uniform schedule does not account for fluctuations in gap size. Adaptive methods, including gap-dependent scheduling or nonlinear interpolation techniques, have the potential to improve performance by decelerating evolution in the vicinity of small gaps. For the sake of experimental simplicity, we utilize $f(s)=s$ in this study.

\end{enumerate}

\section{Algorithm Optimization}
\subsubsection*{3.1 First-Order Approximation of the Discrete Evolution Operator}
In the process of quantum adiabatic evolution, the evolution operator of the system can be expressed in exponential form as $U(t)=e^{-i H t}$. When the time interval $d t$ is extremely small, it can be simplified by using the first-order Taylor expansion approximation as follows:
\begin{equation}
    U(d t) \approx I - i H d t
\end{equation}
where $I$ is the identity matrix. This approximation ignores the higher-order infinitesimals $O\left(d t^2\right)$ and only retains the linear action of the Hamiltonian $H$. For a Hamiltonian $H_s=\left(\begin{array}{cc}0 & B \\ C & 0\end{array}\right)$ with a block structure, the approximate form of the evolution operator is:
\begin{equation}
    e^{-i H d t} \approx\left(\begin{array}{cc}
    I & -i B d t \\
    -i C d t & I
\end{array}\right)
\end{equation}
We can use the block encoding technique \cite{childs2017quantum,camps2024explicit,camps2022fable} for matrix encoding, which can significantly reduce the depth of the quantum circuit (Appendix A).

\subsection{Dynamic Circuit Design Scheme}
During the evolution process of the quantum state, a unique structure of separation between the real and imaginary parts emerges. Taking ${A}$ as a $2\times2$ matrix as an example, we can assume the following:

1. Definition of matrix ${A}$:
\begin{equation}
    {A}=\left(\begin{array}{ll}
    A_{11} & A_{12} \\
    A_{21} & A_{22}
    \end{array}\right)
\end{equation}

2. Definition of matrix ${Q_b}$:
\begin{equation}
    {Q_b}=\left(\begin{array}{ll}
    Q_{11} & Q_{12} \\
    Q_{21} & Q_{22}
    \end{array}\right), \quad {Q_b}=I - |{b}\rangle\langle{b}|
\end{equation}

3. Definition of matrix ${H}$:
\begin{equation}
    \begin{gathered}
{H_0}=\left(\begin{array}{cc}
0 & {Q_b} \\
{Q_b} & 0
\end{array}\right), \quad {H_1}=\left(\begin{array}{cc}
0 & {A}{Q_b} \\
{Q_b}{A} & 0
\end{array}\right) \\
{H_s}=(1 - s){H_0}+s{H_1}=\left(\begin{array}{cc}
0 & {B} \\
{C} & 0
\end{array}\right)
\end{gathered}
\end{equation}

4. Approximation of the Time Evolution Operator
\begin{equation}
    \begin{aligned}
        e^{-i{H}dt} &\approx I - i{H}dt \\
        &= \left(
        \begin{smallmatrix}
            1 & 0 & -i B_{11}dt & -i B_{12}dt \\
            0 & 1 & -i B_{21}dt & -i B_{22}dt \\
            -i C_{11}dt & -i C_{12}dt & 1 & 0 \\
            -i C_{21}dt & -i C_{22}dt & 0 & 1
        \end{smallmatrix}
    \right) 
    \end{aligned}
\end{equation}
Subsequently, we examine the progression of the quantum state as influenced by the time evolution operator. As referenced in the pertinent literature [cite], it is established that the initial condition for the quantum state evolution is $\left(b_1, b_2, 0,0\right)^T$. Our first step involves applying the operator $e^{-i H d t}$ to this initial condition.
\begin{equation}
\begin{aligned}
&\begin{pmatrix}
1 & 0 & -i B_{11} dt & -i B_{12} dt \\
0 & 1 & -i B_{21} dt & -i B_{22} dt \\
-i C_{11} dt & -i C_{12} dt & 1 & 0 \\
-i C_{21} dt & -i C_{22} dt & 0 & 1
\end{pmatrix}
\begin{pmatrix}
b_1 \\
b_2 \\
0 \\
0
\end{pmatrix} \\
&= 
\begin{pmatrix}
b_1 \\
b_2 \\
i(-C_{11} dt \cdot b_1 - C_{12} dt \cdot b_2) \\
i(-C_{21} dt \cdot b_1 - C_{22} dt \cdot b_2)
\end{pmatrix} 
= 
\begin{pmatrix}
x_1^1 \\
x_2^1 \\
i(x_3^1) \\
i(x_4^1)
\end{pmatrix}
\end{aligned}
\end{equation}
The resulting final state retains the same structure. By repeatedly utilizing the time-evolution operator, it is possible to achieve:

\begin{equation}
\prod_{k=1}^L \left(I - i H_k dt\right)
\left(
\begin{smallmatrix}
b_1 \\
b_2 \\
0 \\
0
\end{smallmatrix}
\right)=
\left(
\begin{smallmatrix}
x_1^n \\
x_2^n \\
i\left(x_3^n\right) \\
i\left(x_4^n\right)
\end{smallmatrix}
\right)
\end{equation}

Based on the computational analysis, throughout the quantum discrete adiabatic evolution process, the initial half of the quantum state's components are preserved as real numbers, whereas the latter half consists solely of imaginary numbers. Through the method of mathematical induction, it can be demonstrated that the quantum state consistently retains the following form after the $n$-th step of evolution:

\begin{equation}
    |\psi^{n}\rangle=(x_1^{n}, x_2^{n}, i x_3^{n}, i x_4^{n})^T
\end{equation}

where $x_i^{n} \in \mathbb{R}$. This behavior arises due to the Hamiltonian $H_s$ possessing a non-diagonal block configuration, enabling the independent transmission of real and imaginary parts throughout the evolution.

Quantum measurement is capable of determining the amplitude $\left|x_i\right|^2$ of the components, while the distinction between the real and imaginary components provides direct insight into the phase details: the initial half of the components possess real-valued phases of either 0 or $\pi$, whereas the latter half of the components exhibit imaginary-valued phases of $\pm i$, same as operator $exp(-iHdt)$ (Appendix C). Notably, for an $N\times N$  matrix  A, this technique is generalized to the $2N$-dimensional state,where $\theta_{x_i}$ represents the phase of $x_i$:
\begin{equation}
    \theta_{x_i} = \begin{cases}
0 \text{ or } \pi, & i\in\{1,2,...,N\} \\
\pi/2 \text{ or } 3\pi/2, & i\in\{N+1,N+2,...,2N\}
\end{cases}
\end{equation}
Now measurement techniques have enabled us to acquire both the amplitude and phase information of the intermediate adiabatic evolution state. To fully reconstruct this intermediate state, it is essential to ascertain only the positive or negative sign information of the intermediate adiabatic evolution state.

In this work, we introduce an algorithm for predicting positive and negative signs, grounded in the concept of continuous evolution. By evaluating the alterations in signs of the components between successive evolution steps, we can uniquely ascertain the phase sign. The continuity inherent in adiabatic evolution guarantees that changes in the component signs occur incrementally. We develop a model for sign prediction as follows:

\begin{equation}
    \operatorname{sgn}\left(x_i^{k + 1}\right)=\begin{cases}
\operatorname{sgn}\left(x^{'}_{i}\right), & \left|x_i^{k}\right|<\delta  \\
\operatorname{sgn}\left(x_i^{k}\right), & \text{otherwise} 
\end{cases}
\end{equation}

Let $x^{'}_{i} = 2x_i^{k}-x_i^{k - 1} \approx x^{k+1}_{i}$, where $\delta$ signifies the noise threshold. This algorithm adeptly addresses two scenarios: First, when the component significantly departs from zero, it adopts the sign consistent with the previous iteration. Second, when the component approaches zero, it infers the sign direction by assessing the trend in variation. Empirical assessments demonstrate that this approach efficiently identifies the positive-negative sign information, enabling the integration of mid-circuit measurements (MCM) within the quantum circuit. Consequently, this facilitates the implementation of dynamic circuit designs.

Based on the aforementioned description, given that we possess comprehensive knowledge of the quantum state, we can improve the design of quantum circuits. The conventional quantum circuit (Figure 1(a)) require that each step of discrete adiabatic evolution be connected in series to derive the final solution of the entire quantum discrete adiabatic evolution. However, the refined measurement technique (Figure 1(b)) allows for a thorough measurement immediately following each discrete operation step, thereby yielding a new quantum state. This resultant quantum state is then re-encoded and introduced into the subsequent discrete evolution stage. This stepwise approach considerably diminishes the depth of the quantum circuit involved in computations.

In conventional quantum discrete adiabatic evolution methods\cite{Costa2021OptimalSQ,An2019QuantumLS,Lin2019OptimalPB,Costa2023TheDA}, achieving higher fidelity necessitates a significant extension of the decoherence duration in quantum circuits as precision requirements increase. However, through our proposed quantum discrete adiabatic evolution algorithm design, the improvement of computational accuracy no longer depends on prolonging decoherence time (Figure 2). Remarkably, the decoherence time becomes completely decoupled from precision demands. This advancement substantially reduces the stringent decoherence time requirements for high-fidelity quantum linear systems solving in quantum circuits, thereby unveiling novel possibilities for quantum circuit optimization.

\begin{figure}[htb]
	\centering
	\includegraphics[width=\linewidth]{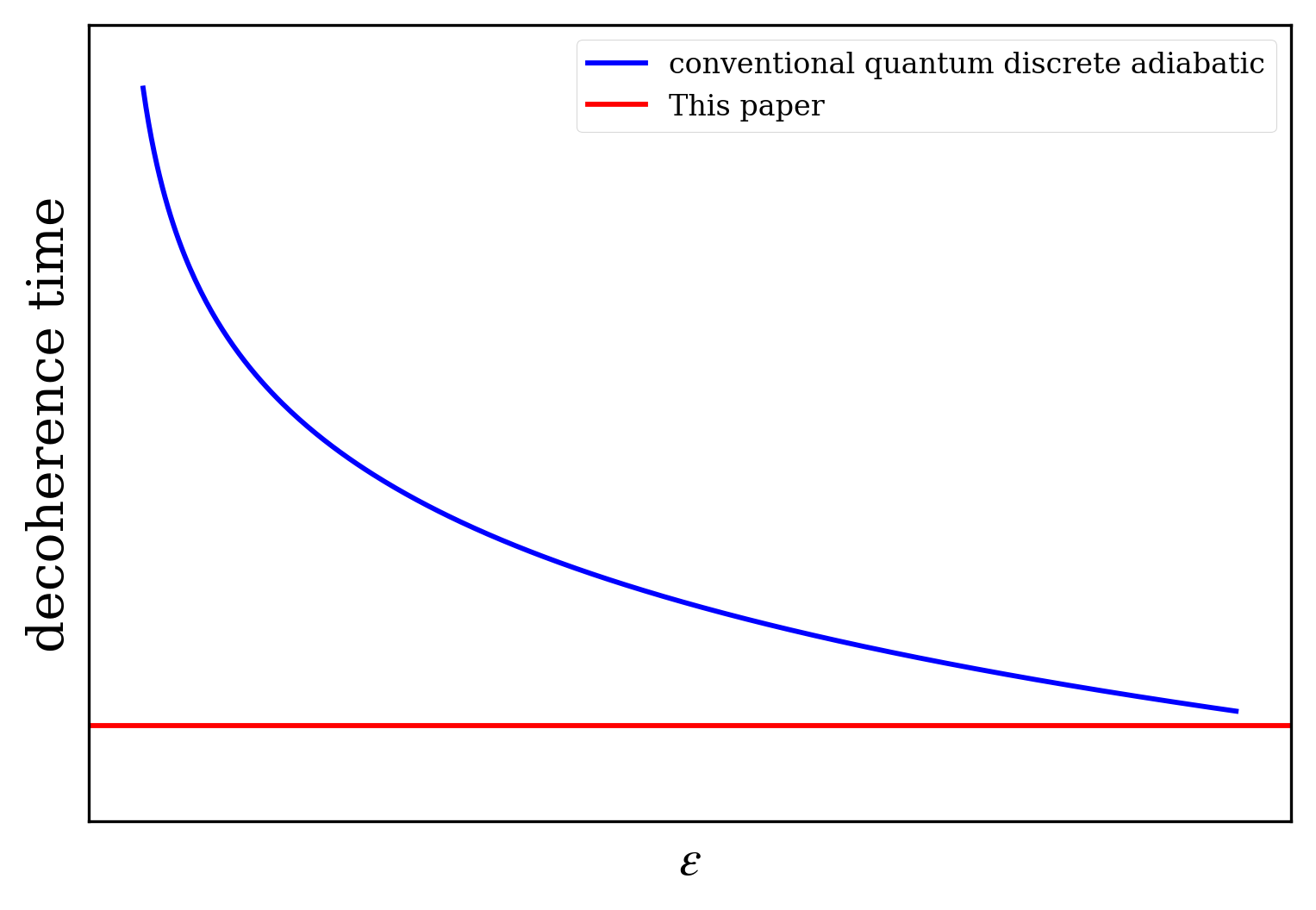}
	\caption{Relationship between Decoherence Time and Precision Requirements
The horizontal axis represents the precision requirement $\epsilon$, and the vertical axis represents the decoherence time. The blue curve represents the conventional quantum discrete adiabatic evolution method, where the decoherence time increases significantly with higher precision requirements.In contrast, the red curve illustrates the quantum discrete adiabatic evolution algorithm proposed in this paper, where the decoherence time is completely decoupled from precision requirements. }
\end{figure}

\subsection{Post-processing Technique}
The amplitudes of the imaginary components $x_{N+1}, x_{N+2}..,, x_{2N}$ in the final state of the evolution reflect the approximation error. Define a truncation threshold $\epsilon = 0.1\|x\|_2$, and implement the post-processing as follows:
\begin{align*}
    &|\tilde{\psi}\rangle = \\ &\begin{cases}
        (x_1, \dots, x_N, 0, \dots, 0)^T, & \text{if } \sum_{j=N+1}^{2N} |x_j|^2 \leq \epsilon^2 \\
        \text{Modify } T, dt, & \text{otherwise}
    \end{cases}
\end{align*}

This technique can concentrate the effective information in the real components and reduce the influence of the imaginary part noise on the solution. For an $N\times N$ system, this technique is generalized to truncate the $2N$-dimensional state to the first $N$ dimensions.
Perform renormalization on the truncated state $|\tilde{\psi}\rangle=(x_1, x_2,.., x_N)^T$:
\begin{equation}
    \left|\psi_{\text {final }}\right\rangle=\frac{1}{\sqrt{x_1^2 +..+ x_N^2}}\begin{pmatrix} x_1 \\ \vdots \\ x_N \end{pmatrix}
\end{equation}
This operation compensates for the normalization loss caused by truncation. Numerical experiments show that after the secondary normalization, the fidelity of the solution is improved.

Define the fidelity between the target state $\left|x_r\right\rangle$ and the calculated state $\left|x_{\text {final }}\right\rangle$ as:
\begin{equation}
    \text { fidelity }=\left|\left\langle x_r \mid x_{\text {final }}\right\rangle\right|
\end{equation}

\begin{figure*}[htb]
	\centering
	\includegraphics[width=\textwidth]{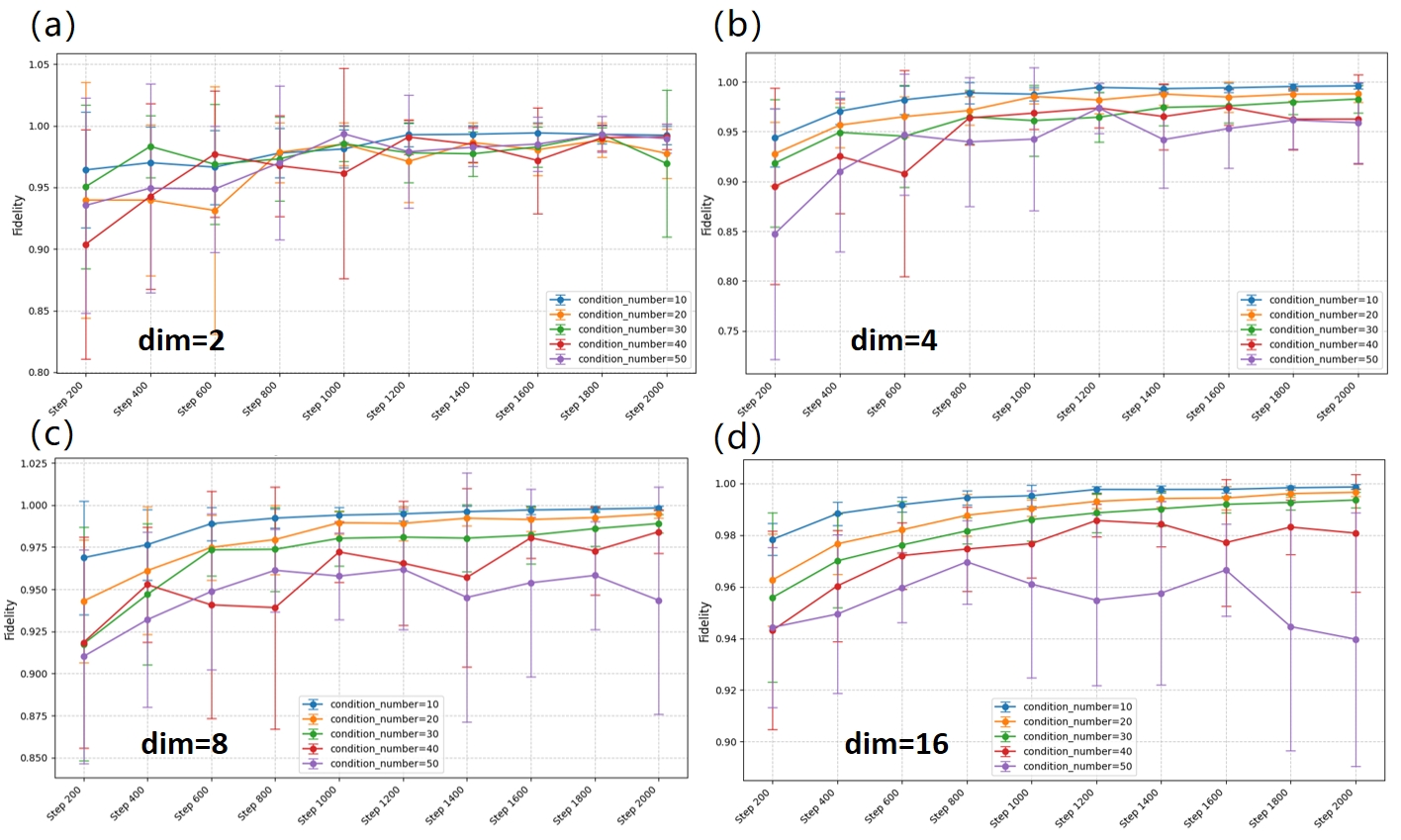}
	\caption{Fidelity Performance Across Different Dimensions and Condition Numbers Without Noise. In the absence of noise, We presents fidelity performance across different matrix dimensions (dim=2, 4, 8, 16) and condition numbers (10, 20, 30, 40, 50) as the number of evolution steps increases from 200 to 2000.}
\end{figure*}

\section{Numerical Tests}

We conducted a systematic experimental investigation on the fidelity characteristics of solving linear systems via a quantum discrete adiabatic evolution algorithm. Figure 3 presents the evolution results for 2×2, 4×4, 8×8, and 16×16 positive definite matrices A with varying condition numbers $(\kappa = 10, 20, 30, 40, 50)$. Each experimental curve comprises 10 data points and their corresponding error ranges. These data points are randomly generated by a program and correspond to suitable matrices A.

The experimental data shows that our discrete adiabatic evolution approach is valid and works well for various matrix sizes. In a noiseless environment, when the number of steps remains constant, the fidelity of solving linear equations decreases as the condition number increases(Figs. 3a-d). 

To enhance the fidelity of solving linear systems, we can increase the number of steps. Under fixed condition numbers, fidelity exhibits a significant positive correlation with the total evolution steps as they increase from 200 to 2000. For small matrices (2×2 and 4×4), fidelity exceeds 95.0\% at 2000 steps (Figs. 3a-b). 

Notably, the algorithm maintains over 90\% fidelity even for the 16×16 large-dimensional matrix system (Fig. 3d), validating the scalability of the proposed scheme. For matrices of fixed dimensions, those with larger condition numbers require more evolution steps to achieve equivalent fidelity. This observation aligns with the theoretically predicted adiabatic time scaling law \cite{Costa2021OptimalSQ,Costa2023TheDA}.

Figure 4 demonstrates the evolution results under a perturbed system with 0.001 decoherence noise. Experiments show that all systems retain over 80\% fidelity after 2000 evolution steps, confirming the algorithm’s robust noise tolerance.

In noisy settings, upping  the number of adiabatic evolution steps and overall time doesn't significantly boost fidelity. More steps mean more quantum circuits and accumulated noise. So, the fidelity - enhancing effect of more steps is offset by noise. Thus, in noisy simulations, increasing steps has a less marked impact on final solution fidelity than in noiseless simulations

It is noteworthy that for stronger noise levels, additional strategies—such as quantum error correction, error mitigation, and optimized evolution step scheduling—are required to sustain high-fidelity performance.

\begin{figure*}[htb]
	\centering
	\includegraphics[width=\textwidth]{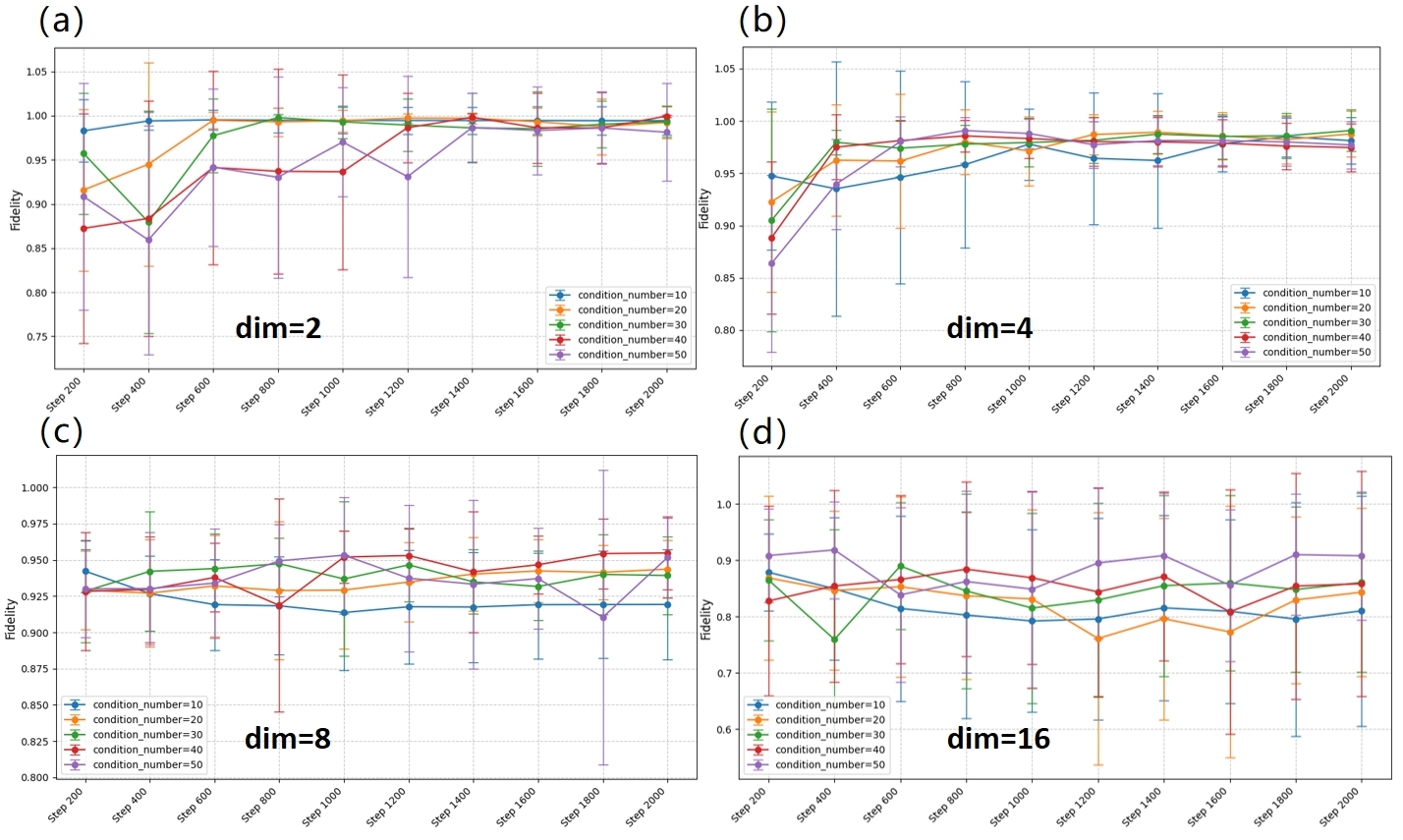}
	\caption{Evolution Results Under a Perturbed System with Decoherence Noise.We presents the evolution results for systems subjected to perturbations with a decoherence noise level of 0.001. For 2×2,4×4,8×8,16×16 matrices, fidelity persists above 80\% at 2000 steps, even under noise, underscoring the algorithm's scalability and noise tolerance.}
\end{figure*}

\section{Conclusion}
The linear scaling of circuit depth with the number of evolution steps in traditional quantum discrete adiabatic approaches poses a critical barrier to their implementation on near-term quantum hardware. To address this challenge, we introduced a co-designed framework that integrates dynamic circuit techniques with real-time classical processing. By reformulating the adiabatic evolution into discrete, dynamically adjustable segments and leveraging circuit multiplexing, our approach reduces the circuit depth scaling from \(O(\text{steps} \times \text{depth}(U))\) to \(O(\text{depth}(U))\), significantly enhancing compatibility with the coherence times of current quantum devices. Key innovations include a first-order approximation of the discrete evolution operator, a dynamic circuit design that exploits real-imaginary component separation, and post-processing techniques to mitigate noise effects.

Benchmarking results demonstrate the robustness of our quantum discrete adiabatic linear solver across systems of dimensions \(W = 2, 4, 8, 16\) and condition numbers \(\kappa = 10\text{--}50\). Under realistic noise models, the solver maintains solution fidelities exceeding \(80\%\) after 2000 evolution steps, showcasing its practicality for noisy intermediate-scale quantum (NISQ) applications. Notably, the algorithm retains scalability to higher dimensions while circumventing challenges such as vanishing gradients in variational methods and the deep circuits required by HHL-based approaches.

This work highlights the potential of hardware-algorithm co-design in unlocking adiabatic quantum computing for practical use cases. Future directions include integrating error mitigation strategies, optimizing adaptive scheduling of evolution steps, and extending the framework to other quantum algorithms requiring time-dependent Hamiltonian simulation. Experimental validation on physical quantum processors and exploration of hybrid quantum-classical workflows for larger-scale problems will further solidify the role of dynamic circuits in advancing quantum linear system solvers.

\bibliography{mybibliography}

\begin{thebibliography}{28}
\providecommand{\natexlab}[1]{#1}
\providecommand{\url}[1]{\texttt{#1}}
\expandafter\ifx\csname urlstyle\endcsname\relax
  \providecommand{\doi}[1]{doi: #1}\else
  \providecommand{\doi}{doi: \begingroup \urlstyle{rm}\Url}\fi

\bibitem[Harrow et~al.(2009)Harrow, Hassidim, and Lloyd]{Harrow2008QuantumAF}
Aram~W Harrow, Avinatan Hassidim, and Seth Lloyd.
\newblock Quantum algorithm for linear systems of equations.
\newblock \emph{Physical review letters}, 103\penalty0 (15):\penalty0 150502,
  2009.

\bibitem[Bravo-Prieto et~al.(2023)Bravo-Prieto, LaRose, Cerezo, Subasi, Cincio,
  and Coles]{BravoPrieto2019VariationalQL}
Carlos Bravo-Prieto, Ryan LaRose, Marco Cerezo, Yigit Subasi, Lukasz Cincio,
  and Patrick~J Coles.
\newblock Variational quantum linear solver.
\newblock \emph{Quantum}, 7:\penalty0 1188, 2023.

\bibitem[Costa et~al.(2022)Costa, An, Sanders, Su, Babbush, and
  Berry]{Costa2021OptimalSQ}
Pedro~CS Costa, Dong An, Yuval~R Sanders, Yuan Su, Ryan Babbush, and Dominic~W
  Berry.
\newblock Optimal scaling quantum linear-systems solver via discrete adiabatic
  theorem.
\newblock \emph{PRX quantum}, 3\penalty0 (4):\penalty0 040303, 2022.

\bibitem[Suba{\c{s}}{\i} et~al.(2019)Suba{\c{s}}{\i}, Somma, and
  Orsucci]{Suba2018QuantumAF}
Yi{\u{g}}it Suba{\c{s}}{\i}, Rolando~D Somma, and Davide Orsucci.
\newblock Quantum algorithms for systems of linear equations inspired by
  adiabatic quantum computing.
\newblock \emph{Physical review letters}, 122\penalty0 (6):\penalty0 060504,
  2019.

\bibitem[An and Lin(2022)]{An2019QuantumLS}
Dong An and Lin Lin.
\newblock Quantum linear system solver based on time-optimal adiabatic quantum
  computing and quantum approximate optimization algorithm.
\newblock \emph{ACM Transactions on Quantum Computing}, 3\penalty0
  (2):\penalty0 1--28, 2022.

\bibitem[Lin and Tong(2020)]{Lin2019OptimalPB}
Lin Lin and Yu~Tong.
\newblock Optimal polynomial based quantum eigenstate filtering with
  application to solving quantum linear systems.
\newblock \emph{Quantum}, 4:\penalty0 361, 2020.

\bibitem[Costa et~al.(2023)Costa, An, Babbush, and Berry]{Costa2023TheDA}
Pedro Costa, Dong An, Ryan Babbush, and Dominic Berry.
\newblock The discrete adiabatic quantum linear system solver has lower
  constant factors than the randomized adiabatic solver.
\newblock 2023.

\bibitem[Cao et~al.(2013)Cao, Papageorgiou, Petras, Traub, and
  Kais]{cao2013quantum}
Yudong Cao, Anargyros Papageorgiou, Iasonas Petras, Joseph Traub, and Sabre
  Kais.
\newblock Quantum algorithm and circuit design solving the poisson equation.
\newblock \emph{New Journal of Physics}, 15\penalty0 (1):\penalty0 013021,
  2013.

\bibitem[Chakraborty et~al.(2018)Chakraborty, Gily{\'e}n, and
  Jeffery]{chakraborty2018power}
Shantanav Chakraborty, Andr{\'a}s Gily{\'e}n, and Stacey Jeffery.
\newblock The power of block-encoded matrix powers: improved regression
  techniques via faster hamiltonian simulation.
\newblock \emph{arXiv preprint arXiv:1804.01973}, 2018.

\bibitem[Childs et~al.(2017)Childs, Kothari, and Somma]{childs2017quantum}
Andrew~M Childs, Robin Kothari, and Rolando~D Somma.
\newblock Quantum algorithm for systems of linear equations with exponentially
  improved dependence on precision.
\newblock \emph{SIAM Journal on Computing}, 46\penalty0 (6):\penalty0
  1920--1950, 2017.

\bibitem[Gily{\'e}n et~al.(2019)Gily{\'e}n, Su, Low, and
  Wiebe]{gilyen2019quantum}
Andr{\'a}s Gily{\'e}n, Yuan Su, Guang~Hao Low, and Nathan Wiebe.
\newblock Quantum singular value transformation and beyond: exponential
  improvements for quantum matrix arithmetics.
\newblock In \emph{Proceedings of the 51st annual ACM SIGACT symposium on
  theory of computing}, pages 193--204, 2019.

\bibitem[Wossnig et~al.(2018)Wossnig, Zhao, and Prakash]{wossnig2018quantum}
Leonard Wossnig, Zhikuan Zhao, and Anupam Prakash.
\newblock Quantum linear system algorithm for dense matrices.
\newblock \emph{Physical review letters}, 120\penalty0 (5):\penalty0 050502,
  2018.

\bibitem[Xu et~al.(2021)Xu, Sun, Endo, Li, Benjamin, and
  Yuan]{xu2021variational}
Xiaosi Xu, Jinzhao Sun, Suguru Endo, Ying Li, Simon~C Benjamin, and Xiao Yuan.
\newblock Variational algorithms for linear algebra.
\newblock \emph{Science Bulletin}, 66\penalty0 (21):\penalty0 2181--2188, 2021.

\bibitem[Jennings et~al.(2024)Jennings, Lostaglio, Lowrie, Pallister, and
  Sornborger]{jennings2024cost}
David Jennings, Matteo Lostaglio, Robert~B Lowrie, Sam Pallister, and Andrew~T
  Sornborger.
\newblock The cost of solving linear differential equations on a quantum
  computer: fast-forwarding to explicit resource counts.
\newblock \emph{Quantum}, 8:\penalty0 1553, 2024.

\bibitem[McClean et~al.(2018)McClean, Boixo, Smelyanskiy, Babbush, and
  Neven]{McClean2018BarrenPI}
Jarrod~R McClean, Sergio Boixo, Vadim~N Smelyanskiy, Ryan Babbush, and Hartmut
  Neven.
\newblock Barren plateaus in quantum neural network training landscapes.
\newblock \emph{Nature communications}, 9\penalty0 (1):\penalty0 4812, 2018.

\bibitem[Cerezo et~al.(2021)Cerezo, Sone, Volkoff, Cincio, and
  Coles]{Cerezo2021CostFD}
Marco Cerezo, Akira Sone, Tyler Volkoff, Lukasz Cincio, and Patrick~J Coles.
\newblock Cost function dependent barren plateaus in shallow parametrized
  quantum circuits.
\newblock \emph{Nature communications}, 12\penalty0 (1):\penalty0 1791, 2021.

\bibitem[Wang et~al.(2021)Wang, Fontana, Cerezo, Sharma, Sone, Cincio, and
  Coles]{Wang2020NoiseinducedBP}
Samson Wang, Enrico Fontana, Marco Cerezo, Kunal Sharma, Akira Sone, Lukasz
  Cincio, and Patrick~J Coles.
\newblock Noise-induced barren plateaus in variational quantum algorithms.
\newblock \emph{Nature communications}, 12\penalty0 (1):\penalty0 6961, 2021.

\bibitem[Ambainis(2012)]{ambainis2012variable}
Andris Ambainis.
\newblock Variable time amplitude amplification and quantum algorithms for
  linear algebra problems.
\newblock In \emph{STACS'12 (29th Symposium on Theoretical Aspects of Computer
  Science)}, volume~14, pages 636--647. LIPIcs, 2012.

\bibitem[C{\'o}rcoles et~al.(2021)C{\'o}rcoles, Takita, Inoue, Lekuch, Minev,
  Chow, and Gambetta]{corcoles2021exploiting}
Antonio~D C{\'o}rcoles, Maika Takita, Ken Inoue, Scott Lekuch, Zlatko~K Minev,
  Jerry~M Chow, and Jay~M Gambetta.
\newblock Exploiting dynamic quantum circuits in a quantum algorithm with
  superconducting qubits.
\newblock \emph{Physical Review Letters}, 127\penalty0 (10):\penalty0 100501,
  2021.

\bibitem[B{\"a}umer et~al.(2024)B{\"a}umer, Tripathi, Wang, Rall, Chen,
  Majumder, Seif, and Minev]{baumer2024efficient}
Elisa B{\"a}umer, Vinay Tripathi, Derek~S Wang, Patrick Rall, Edward~H Chen,
  Swarnadeep Majumder, Alireza Seif, and Zlatko~K Minev.
\newblock Efficient long-range entanglement using dynamic circuits.
\newblock \emph{PRX Quantum}, 5\penalty0 (3):\penalty0 030339, 2024.

\bibitem[Carrera~Vazquez et~al.(2024)Carrera~Vazquez, Tornow, Rist{\`e},
  Woerner, Takita, and Egger]{CarreraVazquez2024CombiningQP}
Almudena Carrera~Vazquez, Caroline Tornow, Diego Rist{\`e}, Stefan Woerner,
  Maika Takita, and Daniel~J Egger.
\newblock Combining quantum processors with real-time classical communication.
\newblock \emph{Nature}, pages 1--5, 2024.

\bibitem[Yu and Fang(2024)]{yu2024symmetry}
Di~Yu and Kun Fang.
\newblock Symmetry-based quantum circuit mapping.
\newblock \emph{Physical Review Applied}, 22\penalty0 (2):\penalty0 024029,
  2024.

\bibitem[Fang et~al.(2023)Fang, Zhang, Shi, and Li]{fang2023dynamic}
Kun Fang, Munan Zhang, Ruqi Shi, and Yinan Li.
\newblock Dynamic quantum circuit compilation.
\newblock \emph{arXiv preprint arXiv:2310.11021}, 2023.

\bibitem[Preskill(2018)]{Preskill2018QuantumCI}
John Preskill.
\newblock Quantum computing in the nisq era and beyond.
\newblock \emph{Quantum}, 2:\penalty0 79, 2018.

\bibitem[Albash and Lidar(2018)]{albash2018adiabatic}
Tameem Albash and Daniel~A Lidar.
\newblock Adiabatic quantum computation.
\newblock \emph{Reviews of Modern Physics}, 90\penalty0 (1):\penalty0 015002,
  2018.

\bibitem[Mori et~al.(2024)Mori, Kawabata, and
  Matsuzaki]{mori2024experimentally}
Yuichiro Mori, Shiro Kawabata, and Yuichiro Matsuzaki.
\newblock How to experimentally evaluate the adiabatic condition for quantum
  annealing.
\newblock \emph{Scientific Reports}, 14\penalty0 (1):\penalty0 8177, 2024.

\bibitem[Camps et~al.(2024)Camps, Lin, Van~Beeumen, and
  Yang]{camps2024explicit}
Daan Camps, Lin Lin, Roel Van~Beeumen, and Chao Yang.
\newblock Explicit quantum circuits for block encodings of certain sparse
  matrices.
\newblock \emph{SIAM Journal on Matrix Analysis and Applications}, 45\penalty0
  (1):\penalty0 801--827, 2024.

\bibitem[Camps and Van~Beeumen(2022)]{camps2022fable}
Daan Camps and Roel Van~Beeumen.
\newblock Fable: Fast approximate quantum circuits for block-encodings.
\newblock pages 104--113, 2022.

\end{thebibliography}

\onecolumn
\newpage % Ensure it's on a new page
\appendix % Switch to appendix mode (A, B, etc. for sections)

\section{Block Encoding for Non-unitary Matrix A}
Since we only consider the first order of Taylor expansion of unitary operator $e^{-iHdt}$, the evolution becomes non-unitary, to implement such non-unitary operations,  here we use the blocking-encoding method.

In quantum computing, we often need to manipulate and transform matrices. However, if these matrices are not unitary, they cannot be directly applied to quantum circuits. The block encoding technique addresses this issue by embedding a non-unitary matrix \( A \) into a larger unitary matrix \( U_A \).

Mathematically, if matrix \( A \) is an \( N \times N \) dimensional matrix, block encoding finds a larger unitary matrix \( U_A \) such that:

\[
U_A =
\begin{bmatrix}
\alpha A & \cdot \\
\cdot & \cdot
\end{bmatrix}
\]

where \( \alpha \) is a scaling factor that ensures \( \alpha A \) has a norm not exceeding 1 (which is required for \( U_A \) to be unitary) and \( \cdot \) represents irrelevant parts. This way, matrix \( A \) can be encoded as a sub-block of the unitary matrix \( U_A \).

The block encoding process requires the multiple auxiliary qubits. Let the initial state \( |in\rangle \) be defined by the auxiliary qubits and the target qubit as follows:

\[
|in\rangle = |0^m\rangle \otimes |v\rangle, \quad |in\rangle =
\begin{pmatrix}
v \\
0
\end{pmatrix}, \quad \|v\| = 1
\]

Applying the block encoding circuit, we obtain:

\[
U_A |in\rangle =
\begin{pmatrix}
\alpha A v \\
\cdot
\end{pmatrix}
= |0^m\rangle (\alpha A |v\rangle) + |0^m\rangle^\perp |\cdot\rangle
\]

The implementation of block encoding \( U_A \) typically involves two important components: Oracle A and Oracle B. Here, we show how these oracles work.

1. Oracle \( O_A \):

\( O_A \) is a black box that provides access to specific data. In block encoding, \( O_A \) is used to construct a quantum circuit capable of accessing and manipulating the elements of matrix \( A \). This Oracle encodes the elements \( A_{i,j} \) of matrix \( A \) into the amplitude of an auxiliary qubit. The operation can be expressed as:

\[
O_A |0\rangle_{\text{anc}} |i\rangle |j\rangle = |A_{i,j}\rangle_{\text{anc}} |i\rangle |j\rangle,
\]

where \( |A_{i,j}\rangle_{\text{anc}} \) represents the state of the auxiliary qubit, defined as:

\[
|A_{i,j}\rangle_{\text{anc}} \equiv A_{i,j} |0\rangle_{\text{anc}} + \sqrt{1 - |A_{i,j}|^2} |1\rangle_{\text{anc}}.
\]

2. Oracle \( O_B \):

\( O_B \) allows an algorithm to query the elements of matrix \( A \). This is typically implemented via a quantum circuit that encodes the indices of the matrix and enables the retrieval of corresponding elements. \( O_B \) correctly maps the matrix index \( (i,j) \). For example, the SWAP gate can be used to exchange indices:

\[
O_B |i\rangle |j\rangle = |j\rangle |i\rangle.
\]

Experimentally, efficiently implementing these Oracles is a crucial challenge. For instance, a series of conditional rotation gates (such as \( Ry \) and \( Rz \) gates) can be used to construct \( O_A \). For matrices with specific structures or sparsity properties, more efficient methods (such as the FABLE technique) \cite{childs2017quantum,camps2024explicit,camps2022fable} can be employed to reduce the required number of gates, thereby optimizing the overall block encoding process.

Next we show how to construct these Oracles using single qubit gates. Suppose the matrix \( A \) is represented as follows, with a corresponding dimension of \( \dim_A = (2^n, 2^n) \), acting on \( n \) qubits. Additionally, the maximum absolute value of \( |A| \) satisfies \( |a_{ij}|_{\max} \leq 1 \).

\[
A =
\begin{pmatrix}
a_{0,0} & \dots & \dots & a_{0,2^n-1} \\
a_{1,0} & \dots & \dots & a_{1,2^n-1} \\
\vdots & \dots & \dots & \vdots \\
a_{(2^n-1),0} & \dots & \dots & a_{(2^n-1),(2^n-1)}
\end{pmatrix}
\]

For any \( k \times k \) matrix \( A \), we can construct an extended matrix \( A' \) by padding rows and columns with zero elements, ensuring its dimension is expanded to \( 2^n \times 2^n \), and then perform block encoding on \( A' \).

\[
A' =
\begin{pmatrix}
A & \dots & 0 \\
\vdots & \dots & \vdots \\
0 & \dots & 0
\end{pmatrix}
\]

Here, we consider only the case where \( \dim_A = (2^n, 2^n) \). Below, we present the quantum circuit for implementing the block encoding of matrix \( A \). The corresponding unitary matrix is \( U_A \), requiring \( n + 1 \) auxiliary qubits, with a total of \( 2n + 1 \) qubits to realize the block encoding.

\[
U_A =
\begin{pmatrix}
\frac{1}{2^n} A & \cdot \\
\cdot & \cdot
\end{pmatrix}
\]

The operator \( O_A \) acts on \( 2n + 1 \) qubits and satisfies the following transformation relationship:

\[
O_A |0\rangle |i\rangle |j\rangle = \left( a_{ij} |0\rangle + \sqrt{1 - |a_{ij}|^2} |1\rangle \right) |i\rangle |j\rangle
\]

where \( i = 0, \dots, 2^n -1 \) and \( j = 0, \dots, 2^n -1 \). The quantum states \( |0\rangle |i\rangle |j\rangle \) correspond to the order from top to bottom. Locally, \( O_A \) is equivalent to the \( R_y(2\theta_{ij}) \) operation, where \( \theta_{ij} = \arccos a_{ij} \), and for \( |1\rangle |i\rangle |j\rangle \), we have:

\[
O_A |1\rangle |i\rangle |j\rangle = \left( a_{ij} |1\rangle + \sqrt{1 - |a_{ij}|^2} |0\rangle \right) |i\rangle |j\rangle
\]

It is easy to see that the construction of \( O_A \) can be achieved using multiple controlled \( R_y \) gates.

\( O_B \) acts on the last \( 2n \) qubits and can be implemented using the SWAP gate:

\[
O_B (|i\rangle |j\rangle) = SWAP(|i\rangle |j\rangle) = |j\rangle |i\rangle
\]

The SWAP operation can be realized using multiple two-qubit swap gates. For example, for a four-qubit case:

\[
SWAP( |2\rangle |3\rangle ) = SWAP( |10\rangle |11\rangle ) = swap_{02} swap_{13} (|10\rangle |11\rangle) = |3\rangle |2\rangle
\]

Thus, the expression for the unitary matrix \( U_A \) is given by:

\[
U_A = (I \otimes H^{\otimes n} \otimes I^{\otimes n}) \cdot (I \otimes SWAP) \cdot O_A \cdot (I \otimes H^{\otimes n} \otimes I^{\otimes n})
\]

Finally we show $U_A$ encodes the elements of matrix A in the top left part. Here, the initial state is set as \( |0\rangle |0\rangle^{\otimes n} |j\rangle \), where \( j = 0, \dots, 2^n - 1 \).
we will use the following convenient relation
\[
|0\rangle |0\rangle^{\otimes n} |j\rangle \xrightarrow{H^{\otimes n}} \frac{1}{\sqrt{2^n}} \sum_{k=0}^{2^n -1} |0\rangle |k\rangle |j\rangle
\]

\[
O_A \xrightarrow{} \frac{1}{\sqrt{2^n}} \sum_{k=0}^{2^n -1} \left( a_{k,j} |0\rangle + \sqrt{1 - |a_{k,j}|^2} |1\rangle \right) |k\rangle |j\rangle
\]

\[
SWAP \xrightarrow{} \frac{1}{\sqrt{2^n}} \sum_{k=0}^{2^n -1} \left( a_{k,j} |0\rangle + \sqrt{1 - |a_{k,j}|^2} |1\rangle \right) |j\rangle |k\rangle \ \ \ (1)
\]

Considering \( i = 0, \dots, 2^n -1 \) and \( j = 0, \dots, 2^n -1 \), the corresponding matrix elements \( (U_A)_{ij} \) are computed as:

\[
(U_A)_{ij} = \langle 0| \langle 0^{\otimes n} | \langle i| U_A (|0\rangle |0^{\otimes n}\rangle |j\rangle )
\]

\[
= \left( \langle 0| \langle 0^{\otimes n} | \langle i| (I \otimes H^{\otimes n} \otimes I^{\otimes n}) (I \otimes SWAP) (O_A) (I \otimes H^{\otimes n} \otimes I^{\otimes n}) (|0\rangle |0^{\otimes n} \rangle |j\rangle ) \right)
\]

The above expression simplifies to:

\[
(U_A)_{ij} = \frac{1}{2^n} \sum_{p=0}^{2^n -1} \sum_{q=0}^{2^n -1} \langle 0| \langle p| \langle i| \left( (a_{qj} |0\rangle + \sqrt{1 - |a_{qj}|^2} |1\rangle ) |j\rangle |q\rangle \right)
\]

\[
= \frac{1}{2^n} \sum_{p=0}^{2^n -1} \sum_{q=0}^{2^n -1} a_{qj} \delta_{pj} \delta_{iq}
\]

\[
= \frac{1}{2^n} a_{ij}
\]

Thus, \( U_A \) encodes \( \frac{1}{2^n} A \) into the top-left block. Moreover, since \( SWAP \), \( H \), and \( I \) gates are all unitary, \( U_A \) is unitary as long as \( O_A \) is unitary.

\section{Quantum State Evolving under Evolution Operator $I-iHdt$}
Here we show the quantum state vector evolution result after applying infinitesimal operator $I-iHdt$.

Setting the initial state $(b_1, b_2, 0, 0)^T$, Acting the evolution operation $I-iHdt$ the first time we obtain
$$
\begin{pmatrix}
1 & 0 & -iB_{11}dt & -iB_{12}dt \\
0 & 1 & -iB_{21}dt & -iB_{22}dt \\
-iC_{11}dt & -iC_{12}dt & 1 & 0 \\
-iC_{21}dt & -iC_{22}dt & 0 & 1
\end{pmatrix}
\begin{pmatrix}
b_1 \\
b_2 \\
0 \\
0
\end{pmatrix}
=
\begin{pmatrix}
b_1 \\
b_2 \\
i(-C_{11}dt \cdot b_1 - C_{12}dt \cdot b_2) \\
i(-C_{21}dt \cdot b_1 - C_{22}dt \cdot b_2)
\end{pmatrix}
=
\begin{pmatrix}
x_1^1 \\
x_2^1 \\
i(x_3^1) \\
i(x_4^1)
\end{pmatrix}
$$
Here $x_1^1,x_2^1,x_3^1,x_4^1$ are real numbers.

Assumed the state vector keeping the form
$(x_1^k, x_2^k, ix_3^k, ix_4^k)^T$ after acting the operator $I-iHdt$  $k$ times,
a step further envolution results the following state vector,
\[
\begin{pmatrix}
1 & 0 & -iB_{11}dt & -iB_{12}dt \\
0 & 1 & -iB_{21}dt & -iB_{22}dt \\
-iC_{11}dt & -iC_{12}dt & 1 & 0 \\
-iC_{21}dt & -iC_{22}dt & 0 & 1
\end{pmatrix}
\begin{pmatrix}
x_1^k \\
x_2^k \\
i(x_3^k) \\
i(x_4^k)
\end{pmatrix}
=
\begin{pmatrix}
x_1^{k+1} \\
x_2^{k+1} \\
i(x_3^{k+1}) \\
i(x_4^{k+1})
\end{pmatrix}
\]
Here we used the induction to show the quantum state vector keeping the form $(x_1^k, x_2^k, ix_3^k, ix_4^k)^T$ unchanged.

\section{Quantum State Evolving under Evolution Operator $exp(-iHdt)$}
Here we show the quantum state vector evolution result after applying infinitesimal operator $exp(-iHdt)$.

Setting the initial state $(b_1, b_2, 0, 0)^T$, Acting the evolution operation $exp(-iHdt)$ the first time we obtain
$$
\begin{pmatrix}
D_{11} & D_{12} & iB_{11} & iB_{12} \\
D_{21} & D_{22} & iB_{21} & iB_{22} \\
iC_{11} & iC_{12} & E_{11} & E_{12} \\
iC_{21} & iC_{22} & E_{21} & E_{22}
\end{pmatrix}
\begin{pmatrix}
b_1 \\
b_2 \\
0 \\
0
\end{pmatrix}
=
\begin{pmatrix}
(D_{11} \cdot b_1 + D_{12} \cdot b_2) \\
(D_{21} \cdot b_1 + D_{22} \cdot b_2) \\
i(C_{11} \cdot b_1 + C_{12} \cdot b_2) \\
i(C_{21} \cdot b_1 + C_{22} \cdot b_2)
\end{pmatrix}
=
\begin{pmatrix}
x_1^1 \\
x_2^1 \\
i(x_3^1) \\
i(x_4^1)
\end{pmatrix}
$$
Here $x_1^1,x_2^1,x_3^1,x_4^1$ are real numbers.

Assumed the state vector keeping the form
$(x_1^k, x_2^k, ix_3^k, ix_4^k)^T$ after acting the operator $exp(-iHdt)$  $k$ times,
a step further envolution results the following state vector,
\[
\begin{pmatrix}
D_{11} & D_{12} & iB_{11} & iB_{12} \\
D_{21} & D_{22} & iB_{21} & iB_{22} \\
iC_{11} & iC_{12} & E_{11} & E_{12} \\
iC_{21} & iC_{22} & E_{21} & E_{22}
\end{pmatrix}
\begin{pmatrix}
x_1^k \\
x_2^k \\
i(x_3^k) \\
i(x_4^k)
\end{pmatrix}
=
\begin{pmatrix}
x_1^{k+1} \\
x_2^{k+1} \\
i(x_3^{k+1}) \\
i(x_4^{k+1})
\end{pmatrix}
\]
Here we used the induction to show the quantum state vector keeping the form $(x_1^k, x_2^k, ix_3^k, ix_4^k)^T$ unchanged.

 % Input the content of the appendix

\clearpage % As suggested in point 1
\end{document}